\documentclass[reprint,superscriptaddress,amssymb,amsmath,aps,prx,longbibliography]{revtex4-2}
\usepackage{minitoc}

\usepackage{graphicx}
\usepackage{dcolumn}
\usepackage{threeparttable}
\usepackage{bm}
\usepackage{siunitx}
\usepackage{booktabs}
\usepackage[usenames,dvipsnames]{xcolor}
\usepackage{tabularx}
\usepackage{array}
\usepackage{colortbl}
\usepackage{braket}
\usepackage{gensymb}
\usepackage{pdfpages}
\usepackage{amsmath}
\usepackage{mathrsfs}
\usepackage{blindtext}
\usepackage{minitoc}
\usepackage{dsfont}
\usepackage{multirow}
\usepackage{xcolor}
\usepackage{soul}
\usepackage{hyperref}

\usepackage{amssymb}

\makeatletter
\patchcmd{\@outputpage@head}{\@ifx{\LS@rot\@undefined}{}{\LS@rot}}{}{}{}
\makeatother

\newcounter{algorithm}

\newcounter{alg}
\renewcommand{\thealg}{\arabic{alg}}

\newenvironment{algorithm}[1][]{%
  \refstepcounter{alg}%
  \par\noindent\textbf{Algorithm~\thealg. #1}\par
  \begin{enumerate}
    \setlength{\itemsep}{0pt}
  }{%
  \end{enumerate}\par
}

\begin{document}

\title{The PID Controller Strikes Back: Classical Controller Helps Mitigate Barren Plateaus in Noisy Variational Quantum Circuits}

\author{Zhehao Yi}
\affiliation{AI, Autonomy, Resilience, Control (AARC) Lab, Electrical \& Computer Engineering, The University of Alabama in Huntsville, AL, 35899, USA}

\author{Rahul Bhadani}
\affiliation{AI, Autonomy, Resilience, Control (AARC) Lab, Electrical \& Computer Engineering, The University of Alabama in Huntsville, AL, 35899, USA}

\date{\today}


\begin{abstract}
Variational quantum algorithms (VQAs) combine the advantages of classical optimization and quantum computation, making them one of the most promising approaches in the Noisy Intermediate-Scale Quantum (NISQ) era. However, when optimized using gradient descent, VQAs often suffer from the vanishing gradient problem, commonly known as the barren plateau. Various methods have been proposed to mitigate this issue. In this work, we propose a hybrid approach that integrates a classical proportional-integral-derivative (PID) controller with a neural network to update the parameters of variational quantum circuits. We refer to this method as NPID, which aims to mitigate the barren plateau. The proposed algorithm is tested on randomly generated quantum input states and random quantum circuits with parametric noise to evaluate its universality, and additional simulations are conducted under different noise rates to examine its robustness. The effectiveness of the proposed method is evaluated based on its convergence speed toward the target cost value. Simulation results show that NPID achieves a convergence efficiency 2–9 times higher than NEQP and QV, with performance fluctuations averaging only 4.45\% across different noise levels. These results highlight the potential of integrating classical control theory into quantum optimization, providing a new perspective for improving the trainability and stability of variational quantum algorithms.
\end{abstract}

\maketitle

\section{Introduction}
Quantum computing has emerged as one of the most promising perspectives in modern computational science \citep{gill2022quantum, meier2025energy, ramezani2020machine, nielsen2010quantum}. Rooted in the principles of quantum mechanics and harnessing the phenomena of entanglement and superposition, quantum computing can perform certain classes of computations more efficiently than its classical counterpart and can represent exponentially richer information states. In quantum computation, the fundamental unit of information is the qubit, which encodes quantum information. With rapid advances in quantum machine learning (QML) and quantum information processing, quantum computing is increasingly being explored across diverse domains such as transportation, healthcare, and finance \citep{herman2023quantum, tayur2024quantum, karahalios2025quantum}.

Quantum machine learning is commonly implemented through variational quantum algorithms (VQAs) \citep{cerezo2021variational}. These algorithms typically consist of four key components: data encoding, parameterized quantum circuits, measurements, and a cost function. The quantum components include the encoding, circuit, and measurement processes, while the cost function represents the classical optimization part. This hybrid structure closely resembles that of traditional neural networks, where the cost function guides the parameter updates throughout the model. Owing to this flexible architecture, VQAs have become one of the most promising frameworks for quantum machine learning in the Noisy Intermediate-Scale Quantum (NISQ) era \citep{lau2022nisq}.

Currently, a significant portion of research in quantum machine learning relies on VQAs. In the transportation domain, researchers have employed variational quantum circuits (VQCs) to construct Quantum Deep Convolutional Neural Networks (QDCNNs) for modeling shadow image processing under complex environmental conditions \citep{meghanath2025qdcnn}, achieving promising performance. Other studies have proposed the $UU^{\dagger}$ algorithm for quantum image processing, demonstrating robustness and strong resistance to noise \citep{innan2025qnnvrcs}. A hybrid quantum genetic algorithm (HQGA) has also been developed for structural damage identification, improving detection efficiency while maintaining stability \citep{xu2025hybrid}. In addition, optimization-oriented approaches such as the Quantum Approximate Optimization Algorithm (QAOA) \citep{zhou2023qaoa} and the Quantum Alternating Direction Method of Multipliers (QADMM) \citep{nie2025qadmm} have been explored for target optimization and distributed quantum computing, respectively.

Despite a large number of advantages of VQAs, they suffer from a critical problem known as the barren plateau \citep{larocca2025barren, mcclean2018barren}. This issue arises when the number of qubits or the depth of the quantum circuit increases. As a result, the gradient used to update model parameters vanishes exponentially. The optimization process often becomes trapped in local optima or stalls entirely. This can lead to significant computational overhead without any meaningful improvement in model performance. The barren plateau effect is influenced by various factors, including circuit initialization, depth, cost function design, and quantum noise. To address this problem, several mitigation strategies have been proposed. These include tailored cost function design \citep{cerezo2021cost}, layerwise training \citep{skolik2021layerwise}, improved parameter initialization \citep{kashif2024alleviating}, neural network generation of quantum states \citep{yi2025neural}, and the use of neural-network-based optimizers for parameter updates \citep{yi2025enhancing, friedrich2022avoiding}.

In this paper, we propose an alternative approach to mitigate the barren plateau problem. We employ a neural proportional–integral–derivative (NPID) controller to update the parameters of a variational quantum algorithm and find that this classical control mechanism effectively enhances the performance of the VQAs under noise.

This article is organized as follows. Section \ref{sec2} provides the necessary background information. Section \ref{sec3} describes the proposed methodology. Section \ref{sec4} presents and analyzes the experimental results, and Section \ref{sec5} concludes the paper with a summary and future outlook.

\section{Background}\label{sec2}
\subsection{Variational Quantum Algorithms}
\begin{figure}[htbp]
    \centering
    \includegraphics[width=0.65\linewidth]{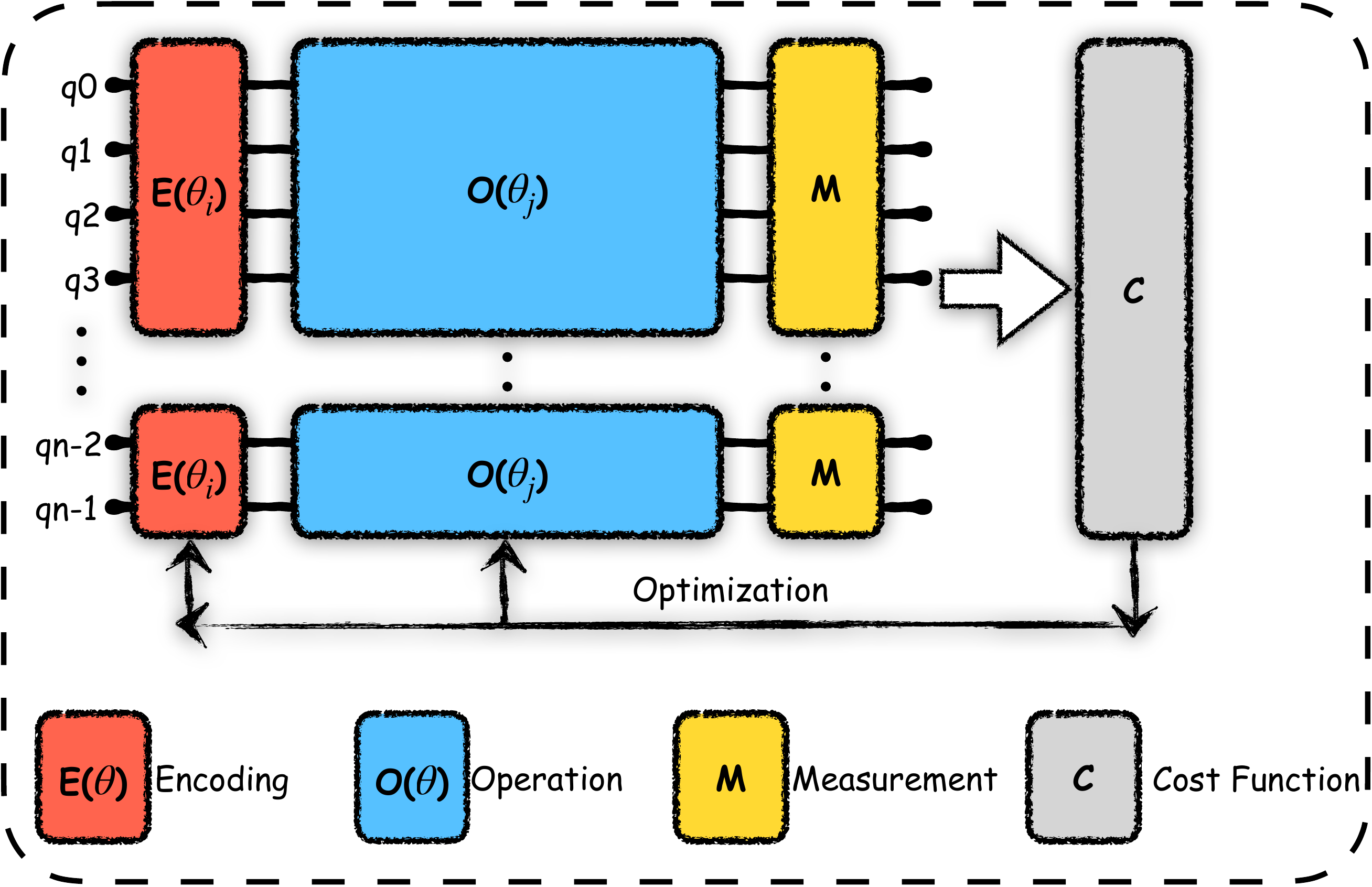}
    \caption{The construction of the Variational Quantum Algorithm.}
    \label{fig: vqc}
    \vspace{-10pt}
\end{figure}

Variational quantum algorithms are hybrid quantum–classical frameworks that optimize parameterized quantum circuits to minimize a given cost function. A VQA consists of four important components: encoding, quantum operation, measurement, and classical optimization. In the encoding stage, classical data are transformed into quantum states by using angle, basis, or amplitude encoding to serve as inputs to the circuit \citep{rath2024quantum}, as shown in Figure \ref{fig: vqc}.

The quantum circuit is composed of parameterized single-qubit rotation gates (e.g., $Rx$, $Ry$, $Rz$) and multi-qubit entangling gates (e.g., $CNOT$, $CZ$), which together define a tunable operation on the quantum state \citep{wong2023quantum, williams2011quantum, divincenzo1998quantum}. During this transformation process, intermediate measurements are prohibited because they would collapse the superposition, halting further computation.

After the operation completes, measurement yields a probability distribution over computational basis states. These probabilities are used to evaluate the cost function, whose gradient information is used in the classical parameter optimization. While VQAs have demonstrated strong expressivity and adaptability across quantum machine learning tasks, they also suffer from the barren plateau problem, in which gradients vanish exponentially as the number of qubits or circuit depth increases.

\subsection{Barren Plateau}
Barren plateaus are among the major challenges hindering the scalability of VQAs \citep{friedrich2022avoiding, mcclean2018barren}. Since VQAs typically rely on gradient-based optimization to update circuit parameters, the occurrence of a barren plateau causes the gradients to vanish, which, as they approach zero, impedes the training process. Let the parameterized quantum circuit be represented by $U(\theta_i)$, the input quantum state is $\ket{\psi_{in}}$, then the output quantum state is:
\begin{equation}
    \ket{\psi_{out}} = U(\theta_i)\ket{\psi_{in}}
    \label{eq:outstate}
\end{equation}
Given a cost function $\mathcal{L}$, the gradient of $\mathcal{L}$ with respect to each trainable parameters $\theta_i$, can be expressed as:
\begin{equation}
    \frac{\partial \mathcal{L}}{\partial \theta_i} = \frac{\partial \bra{\psi_{out}}\hat{M}\ket{\psi_{out}}}{\partial \theta_i}
    \label{eq:partialcost}
\end{equation}
where the $\hat{M}$ denotes the measurement consequences~\citep{cerezo2021variational}.

As shown in previous analyses, in deep, highly entangled circuits or those containing a large number of qubits, the gradient converges exponentially toward zero, as expressed in Equation \eqref{eq: bp}, 
\begin{equation}
    \frac{\partial \mathcal{L}}{\partial \theta_i} \leq G(n), G(n)\in O(\frac{1}{a^n}), a\geq1
    \label{eq: bp}
\end{equation}
rendering the optimization ineffective and resulting in the barren plateau.

\subsection{Proportional–integral–derivative Controller}
Proportional–integral–derivative controllers have been widely employed in industrial and engineering systems for decades \citep{astrom1995pid}. In a closed-loop linear control system, the controller regulates the system output to track a desired reference. The error signal $e(t)$, defined as the difference between the reference and the actual output, serves as the basis for adjustment.
The PID control law comprises three components: Proportional control multiplies the error by a proportional gain $K_p$:
\begin{equation}
    P = K_pe(t)
\end{equation}
Integral control integrates the error over time and is scaled by an integral $K_i$:
\begin{equation}
    I = K_i\int_{0}^{t}e(\tau)d\tau
\end{equation}
and Derivative control differentiates the error with respect to time and multiplies it by a derivative gain $K_d$:
\begin{equation}
    D = K_d \frac{de(t)}{dt}
\end{equation}

By combining these three terms, the PID controller uses current, accumulated, and predicted error information to tune the control system, thereby enhancing system stability. However, specific control modes can be realized by setting one or more of the gains to zero, resulting in P, PI, or PD controllers.

\section{Method}
\label{sec3}
In this section, we present the proposed methodology. Since the PID controller is fundamentally designed for linear systems \citep{astrom1995pid}, we first demonstrate in Appendix \ref{secA1} that a parameterized quantum circuit can be regarded as a linear system under certain conditions.

To ensure the generality of the experiments, random quantum input states were employed. These states were generated by sequentially applying quantum rotation gates to the ground state $\ket{0}$, shown in Equation \eqref{eq: rdinputsate}, with the rotation parameters randomly initialized to guarantee the randomness of the input states.
\begin{equation}
    \ket{\psi_{in}} = \bigotimes_1^{n}\{R_z(\theta_k)R_y(\theta_j)R_x(\theta_i)\ket{0}\}
    \label{eq: rdinputsate}
\end{equation}

\begin{figure}
    \centering
    \includegraphics[width=0.9\linewidth]{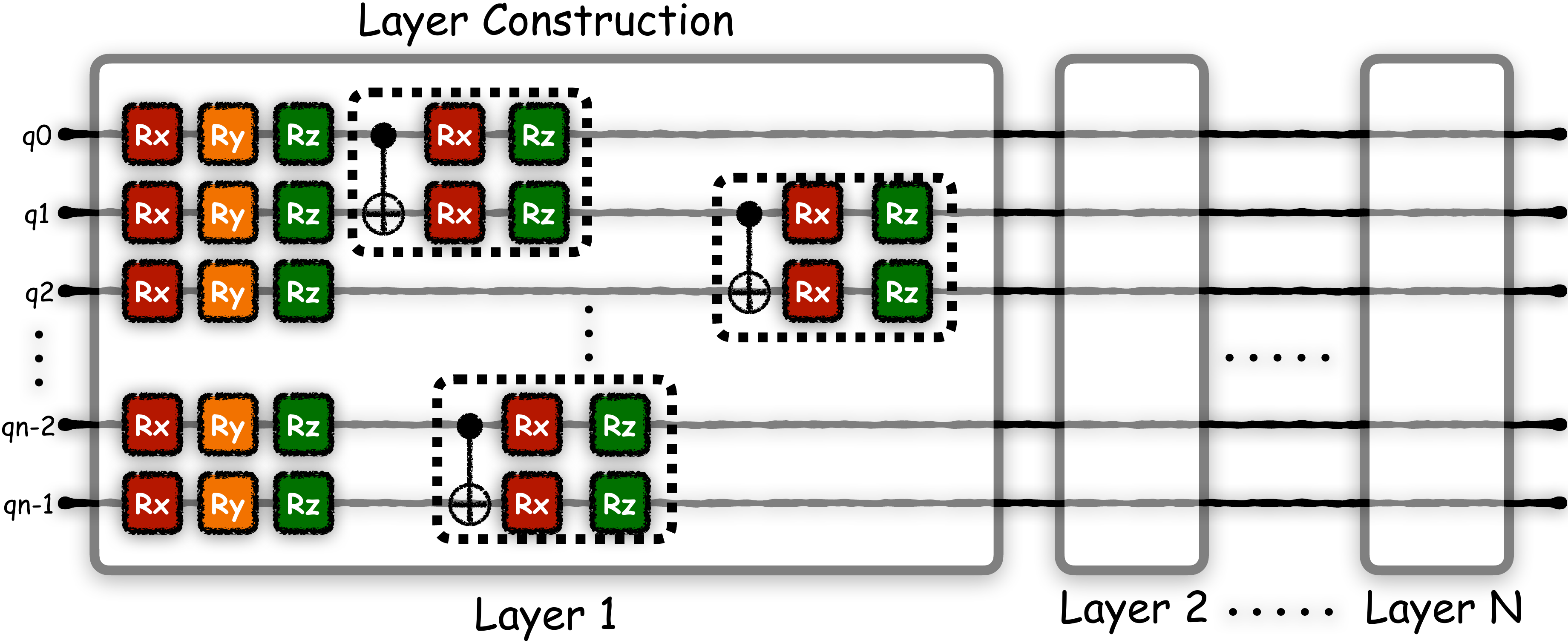}
    \caption{The Construction of the random quantum circuit.
The gray boxes represent individual circuit layers, while the black dashed boxes indicate the gates applied between pairs of randomly selected qubits.}
    \label{fig: rdqc}
    \vspace{-24pt}
\end{figure}

We constructed random quantum circuits with circuit depth $D$, representing the number of layers. Each layer begin with single-qubit rotation gates applied to all qubits. Qubits were then randomly paired and entangled using CNOT gates, followed by additional $R_x$ and $R_z$ rotation gats to increase complexity, as shown in Figure \ref{fig: rdqc}. Because true randomness cannot be achieved computationally, we maximized the circuits complexity to approximate the behavior of truly random quantum circuits. Given that practical computation is never perfectly noise-free, we introduced parameter noise was to all parameterized gates. For any parameters set $\theta$, a random perturbation $\Delta$ was added; $\delta$ represents the noise rate, and $\alpha$ is a random parameter set of the same as $\theta$, as shown in the following Equation \eqref{eq: nosie}.
\begin{equation}
    \begin{aligned}
        \hat{\theta} &= \theta + \Delta \\
        \Delta &= \delta \cdot \alpha
    \end{aligned}
    \label{eq: nosie}
\end{equation}

Next, we employed the cost function in Equation \eqref{eq: costfunction} to compute the loss value, consistent with that used in previous studies \citep{friedrich2022avoiding}:
\begin{equation}
    \mathcal{L} = 1- \frac{1}{n}\sum_{i1}^{n}Tr\{ ( \ket{0}\bra{0}_i \otimes I_{\hat{i}} ) \ket{\psi_{in}} \}
    \label{eq: costfunction}
\end{equation}
In this formulation, the $i$-th qubit interacts with 
$\ket{\psi_{in}}$, while $I_{\hat{i}}$ interacts with all other qubits except $i$. This formula represents one minus the probability that each qubit in the output quantum state is in the ground state $\ket{0}$. Minimizing this cost function requires the latter term to approach 1, which in turn enforces the number of output qubits to be close to $\ket{0}^{\otimes n}$.

After obtaining the result from the cost function $e = \mathcal{L}$, the PID control process begins. First, the three components of the PID controller are computed. In the proportional control term, $P_e$ is defined as the current output loss $e$. In the integral control term, $I_e$ is defined as the sum of the current and previous loss $e_{pre}$, approximating the integration over time. In the derivative control term, $D_e$ is defined as the difference between the current and previous loss. Second, the three PID coefficients are determined.
Unlike traditional PID control, where the coefficients are fixed, we introduce a small fully connected neural network that dynamically adjusts these coefficients during control. This neural network receives $P_e$, $I_e$, $D_e$ and $e$ as input and the outputs the three adaptive coefficients $K_p$, $K_i$, and $P_d$. Next, the corresponding terms are multiplied and summed to obtain the PID output $O_{pid}$. The gradient with respect to 
parameters set $\frac{\partial C}{\partial \theta}$ is then computed, and the parameter update process for the list 
$\theta$ is expressed as Equation \eqref{eq: updatetheta}, where $\hat{\theta}$ means the noisy parameter list, and $lr$ means the learning rate.
\begin{equation}
    \hat{\theta}_{i+1} =\hat{\theta_i} - lr \cdot \frac{\partial \mathcal{L}}{\partial \theta} \cdot O_{pid}
    \label{eq: updatetheta}
\end{equation}
The algorithm is as follows:

\begin{algorithm}[Calculate update parameters set $\hat{\theta}_{i+1}$]\label{algo1}
  \item $e$ is the current loss, $e_{pre}$ is the previous loss, $\frac{\partial \mathcal{L}}{\partial \theta}$ is the gradient, $lr$ means the learning rate, $\hat{\theta_i}$ means the previous parameters set, and $\phi$ denotes the neural network.
  \item $P_e = e$
  \item $I_e = e + e_{\text{pre}}$
  \item $D_e = e - e_{\text{pre}}$
  \item $K_p, K_i, K_d = \phi(e, P_e, I_e, D_e)$
  \item $O_{pid} = K_p P_e + K_i I_e + K_d D_e$
  \item $\hat{\theta}_{i+1} = \hat{\theta}_i - lr \cdot 
        \dfrac{\partial \mathcal{L}}{\partial \theta} \cdot O_{pid}$
\end{algorithm}

\begin{figure}[htbp]
    \centering
    \includegraphics[width=0.9\linewidth]{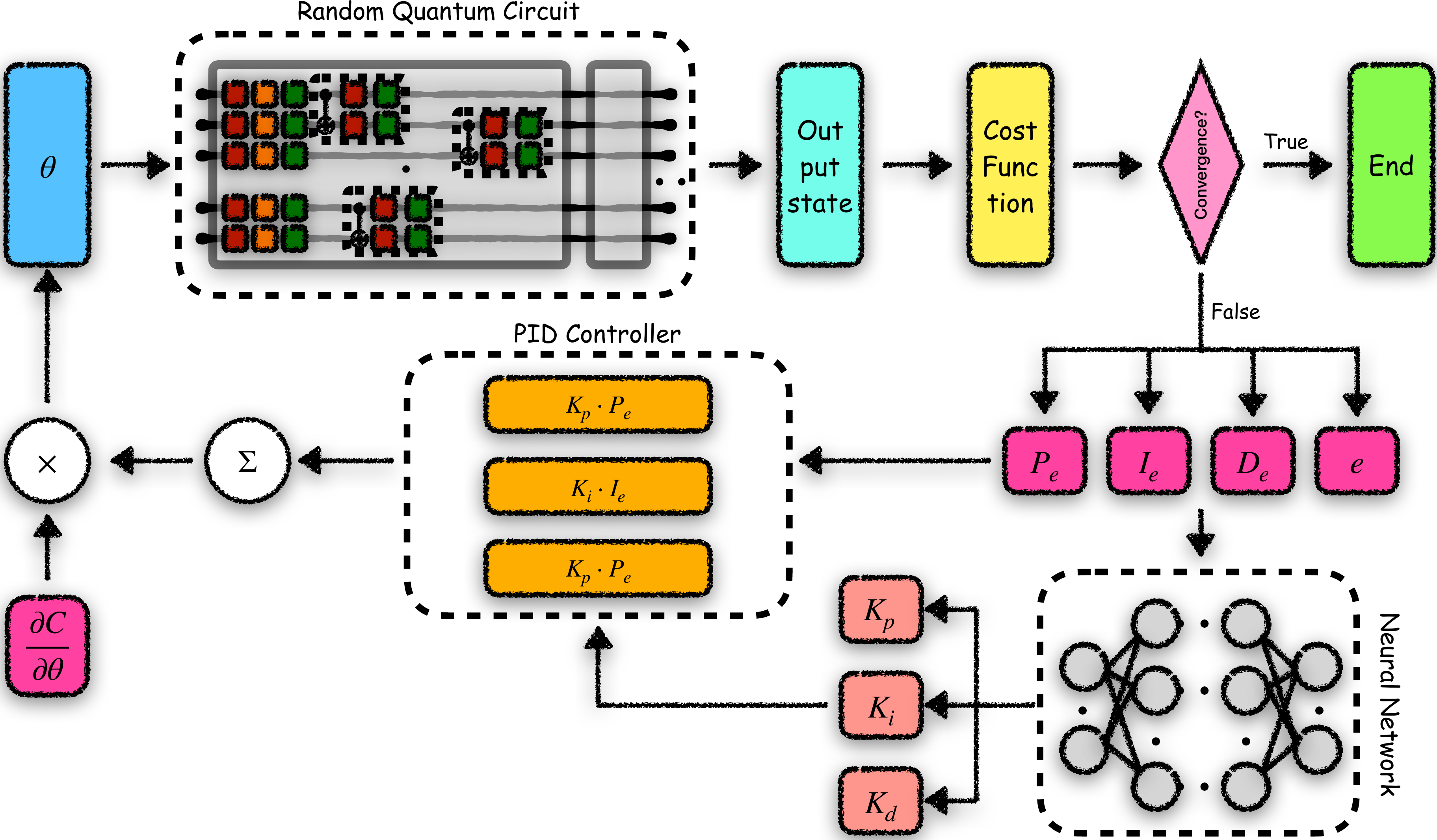}
    \caption{The construction of the Neural Proportional–Integral–Derivative Model.}
    \label{fig: flowchatNPID}
\end{figure}

The overall training flowchart is shown in Figure \ref{fig: flowchatNPID}. NPID denotes the Neural PID Model, NEQP represents the Neural Enhanced Quantum Parametric Model, and QV refers to the Standard Quantum Vanilla Model. In NEQP, the neural network receives a random parameter set as input and outputs a parameter vector matching the required circuit dimensionality \citep{friedrich2022avoiding, yi2025enhancing}. These parameters are then sequentially applied to the quantum circuit to generate the output state. The cost function is computed based on this output state, and the resulting loss value is used to update the weights $w$ and biases $b$ of the neural network, rather than the input parameters themselves. In NPID, the computed loss value $e$ serves two purposes: it is used in the PID control computation to update the parameter list $\theta$ of the quantum circuit, and simultaneously drives the gradient updates of the neural network that generates the PID gains. In contrast, QV directly generates the required parameter vector and updates it through the same gradient-based optimization steps.

The architectures of the neural networks used in the simulations are shown in the Table \ref{tab: constnn}. NPID and NEQP share the same number of layers, while NPID and NEQP-S have an identical number of neurons per layer except for the final output, and NEQP-L employs a larger number of neurons in each layer. A Softplus activation function is added at the output of the NPID network, ensuring that the PID controller gains remain positive, as is typical in control systems \citep{ogata2010modern, dorf2011modern}. The goal of maintaining an identical number of layers while varying the neuron count enables an evaluation of the proposed model’s learning efficiency \citep{lu2017expressive}.
\begin{table}[h]
 \centering \caption{The construction of neural networks}
 \begin{threeparttable}
\begin{tabular}{ccc}
    \hline
    NPID & NEQP-S &NEQP-L \\
    \hline
    Linear(4, 32) & Linear(4, 32) & Linear(32, 256) \\
    Tanh() & Tanh() & Tanh() \\
    Linear(32, 64) & Linear(32, 64)& Linear(256, 256) \\
    Tanh() & Tanh() & Tanh() \\
    Linear(64, 3) & Linear(64, $\theta$\tnote{*}) & Linear(256, $\theta$\tnote{*}) \\
    Softplus() & \\
    \hline
   \end{tabular}
   \begin{tablenotes}
       \footnotesize
       \item[*]The parameters list for the quantum circuit. 
   \end{tablenotes}
   \label{construction}
   \end{threeparttable}
   \label{tab: constnn}
\end{table}

In summary, we propose a hybrid classical control system with a quantum variational framework to mitigate the barren plateau. In the following section, we will present our simulation results.

\section{Results}
\label{sec4}
We first describe the experimental setup.
The number of qubits ranges from 7 to 12. The depth of each random quantum circuit is set to $D = n^2\log n$, where $n$ denotes the number of qubits. For each of the four models, we conduct five independent runs with distinct circuit initializations and parameter seeds. All models are optimized using stochastic gradient descent (SGD). The training objective is to reduce the cost function value $\mathcal{L}$ below 0.001, and the noise rate is 0.01. All models are developed in PyTorch \citep{paszke2019pytorch}, and the quantum circuits are constructed with TorchQuantum \citep{hanruiwang2022quantumnas}. The experiments are executed on a workstation equipped with an AMD Ryzen 9 7960X 24-core processor and an NVIDIA RTX 4090 GPU with 24 GB of memory.

\begin{figure*}[htbp]
    \centering

    \begin{minipage}[b]{0.48\textwidth}
        \centering
        \includegraphics[width=\textwidth]{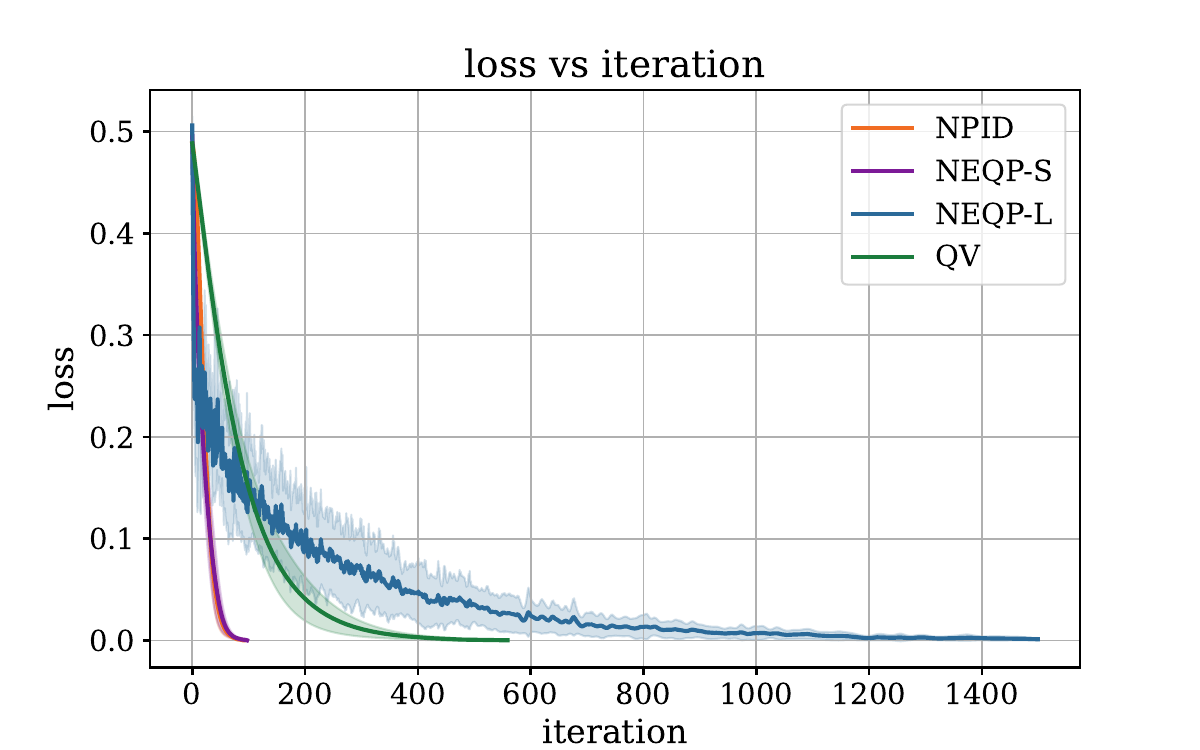}
        {\footnotesize (a) 7 Qubits}
    \end{minipage}
    \hfill
    \begin{minipage}[b]{0.48\textwidth}
        \centering
        \includegraphics[width=\textwidth]{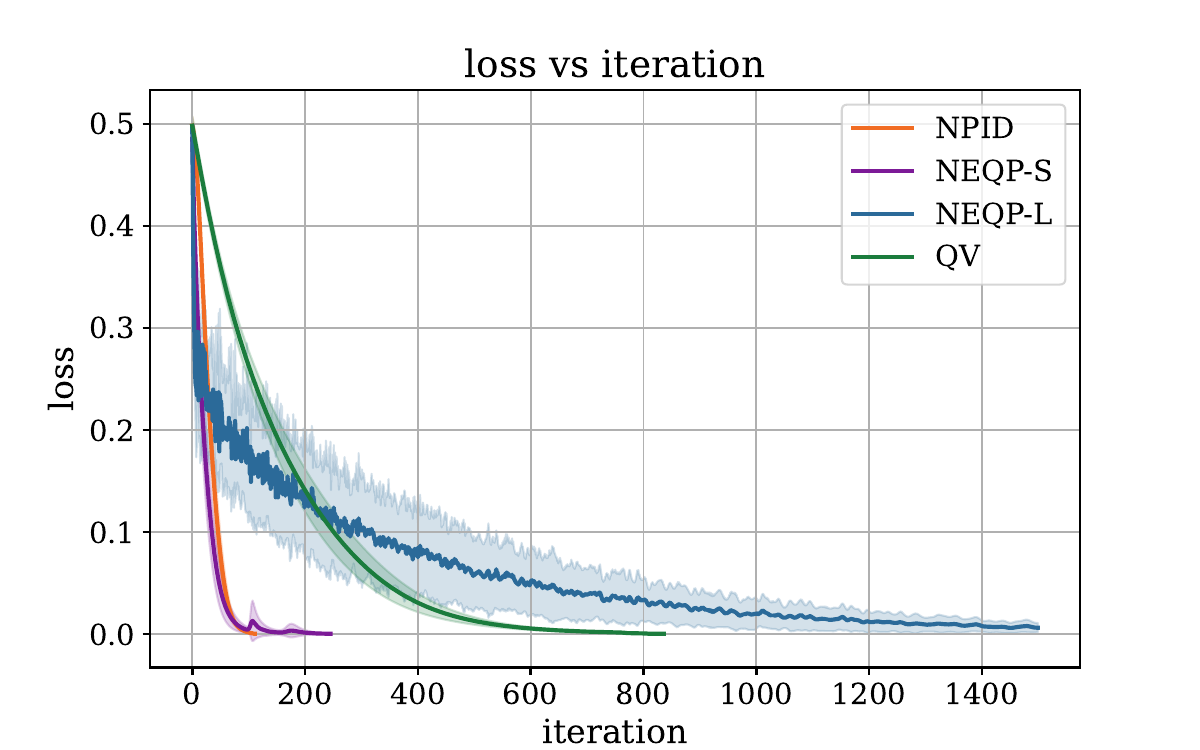}
        {\footnotesize (b) 8 Qubits}
    \end{minipage}

    \vspace{4pt}

    \begin{minipage}[b]{0.48\textwidth}
        \centering
        \includegraphics[width=\textwidth]{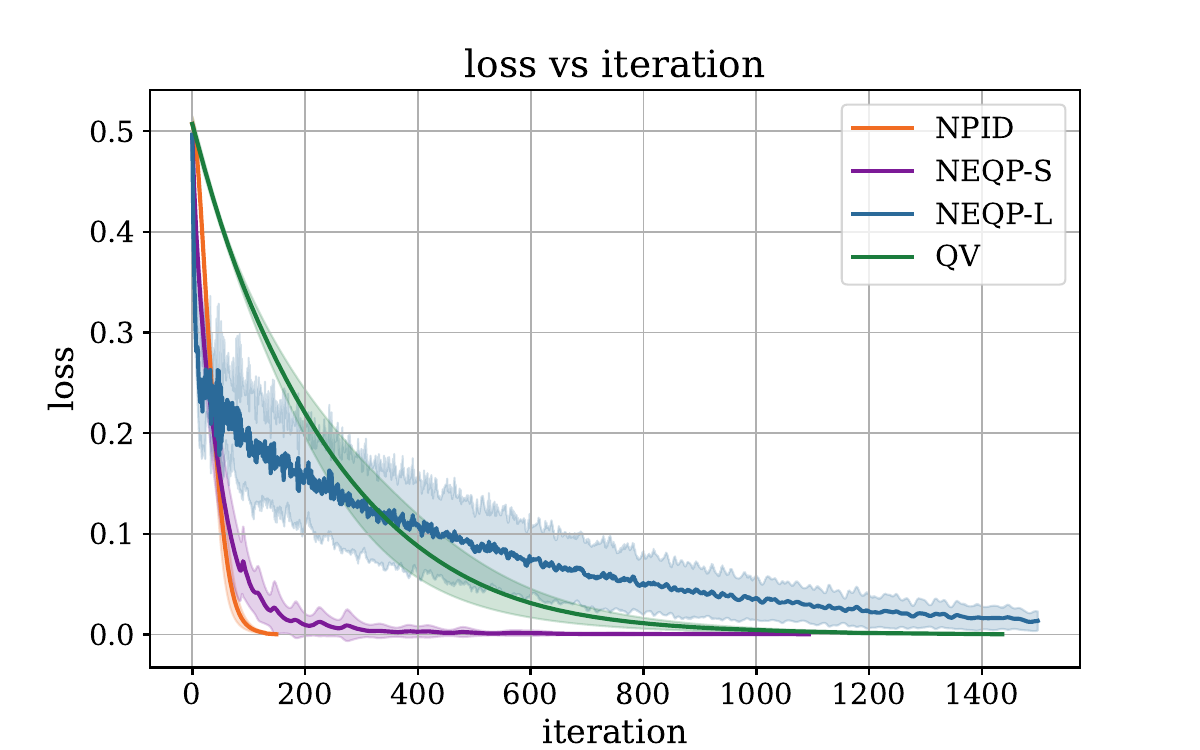}
        {\footnotesize (c) 9 Qubits}
    \end{minipage}
    \hfill
    \begin{minipage}[b]{0.48\textwidth}
        \centering
        \includegraphics[width=\textwidth]{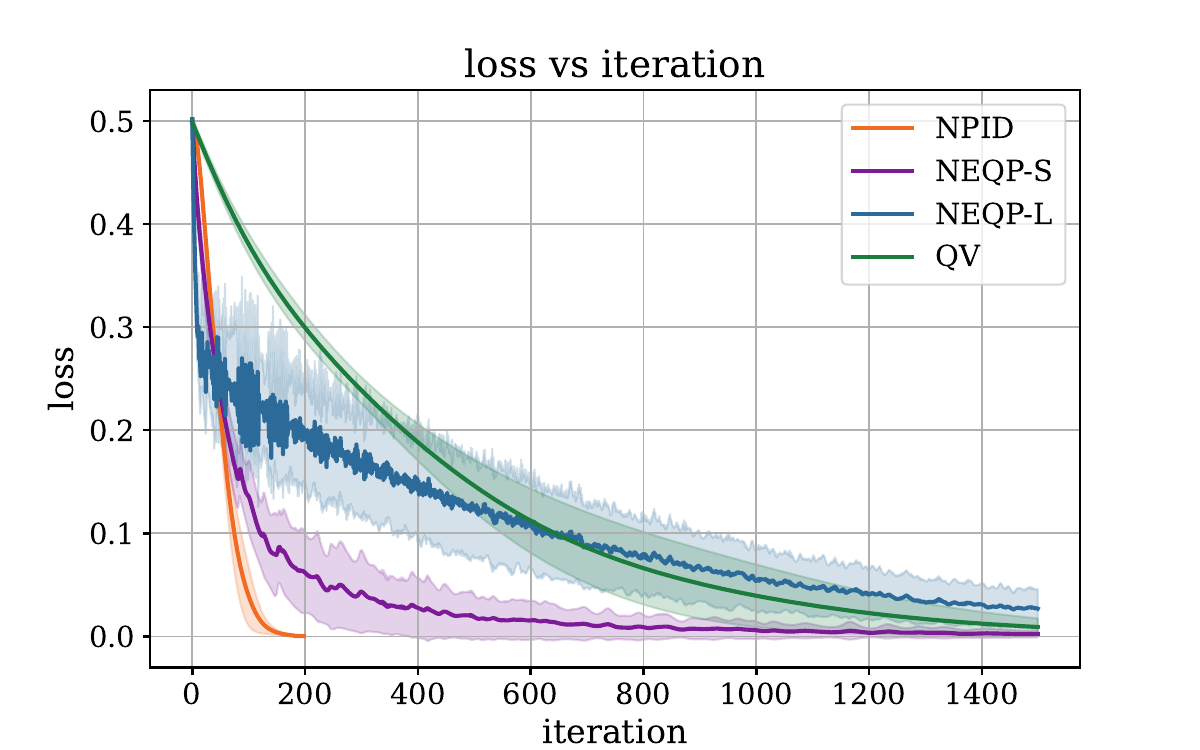}
        {\footnotesize (d) 10 Qubits}
    \end{minipage}

    \vspace{4pt}

    \begin{minipage}[b]{0.48\textwidth}
        \centering
        \includegraphics[width=\textwidth]{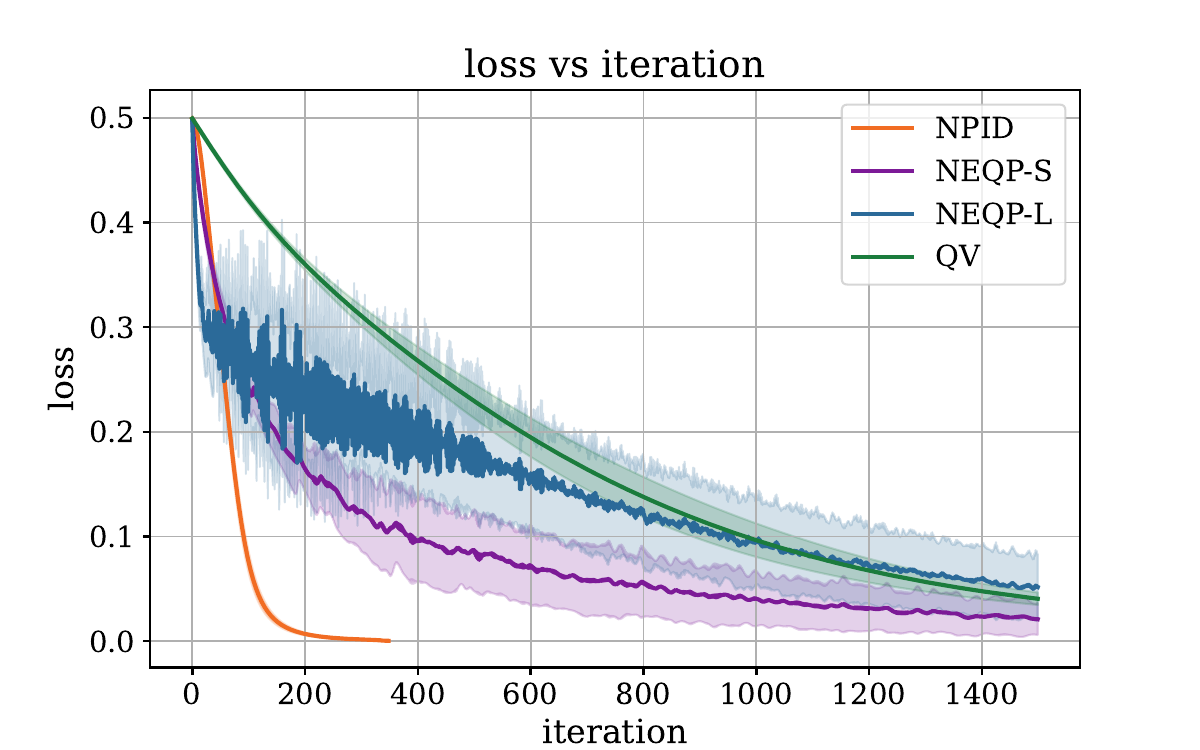}
        {\footnotesize (e) 11 Qubits}
    \end{minipage}
    \hfill
    \begin{minipage}[b]{0.48\textwidth}
        \centering
        \includegraphics[width=\textwidth]{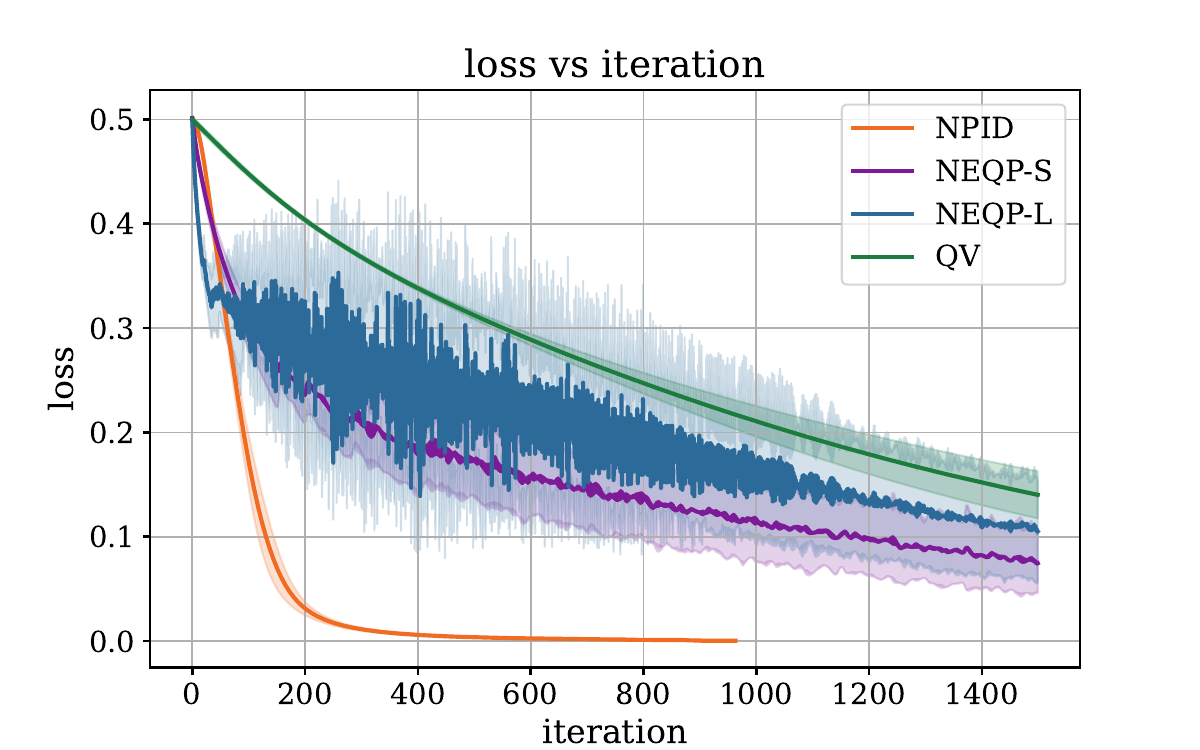}
        {\footnotesize (f) 12 Qubits}
    \end{minipage}

    \caption{Average loss versus number of iterations.
    The shaded regions represent the mean $\pm$ standard deviation over five independent runs.}
    \label{fig: lossvsiter}
\end{figure*}

Figure \ref{fig: lossvsiter} illustrates the relationship between the number of iterations and the loss value. In this experiment, the maximum number of iterations was set to 1500. As shown in the figure, as the number of qubits increases, NEQP-L consistently requires the full number of iterations to converge, while the iterations needed by NEQP-S and QV also grow rapidly with system size. In contrast, NPID maintains a consistently fast decrease in loss. Although its iteration count increases slightly with the number of qubits, it remains significantly lower than that of the other two models. This demonstrates that NPID effectively mitigates the barren plateau effect associated with increasing qubit count and circuit depth. The shaded regions represent the mean and variance of the recorded loss values, showing that NPID exhibits stable convergence behavior. In contrast, QV and NEQP-S perform worse as the number of qubits increases, while NEQP-S and NEQP-L display pronounced fluctuations under noisy quantum circuits and remain trapped in a barren plateau. These results also indicate that simply increasing the number of neurons in a neural network does not necessarily improve its performance. Although NEQP-S performs well for smaller qubits, it, like NEQP-L, eventually encounters the barren plateau as the system size increases.

\begin{figure}
    \centering
    \includegraphics[width=0.8\linewidth]{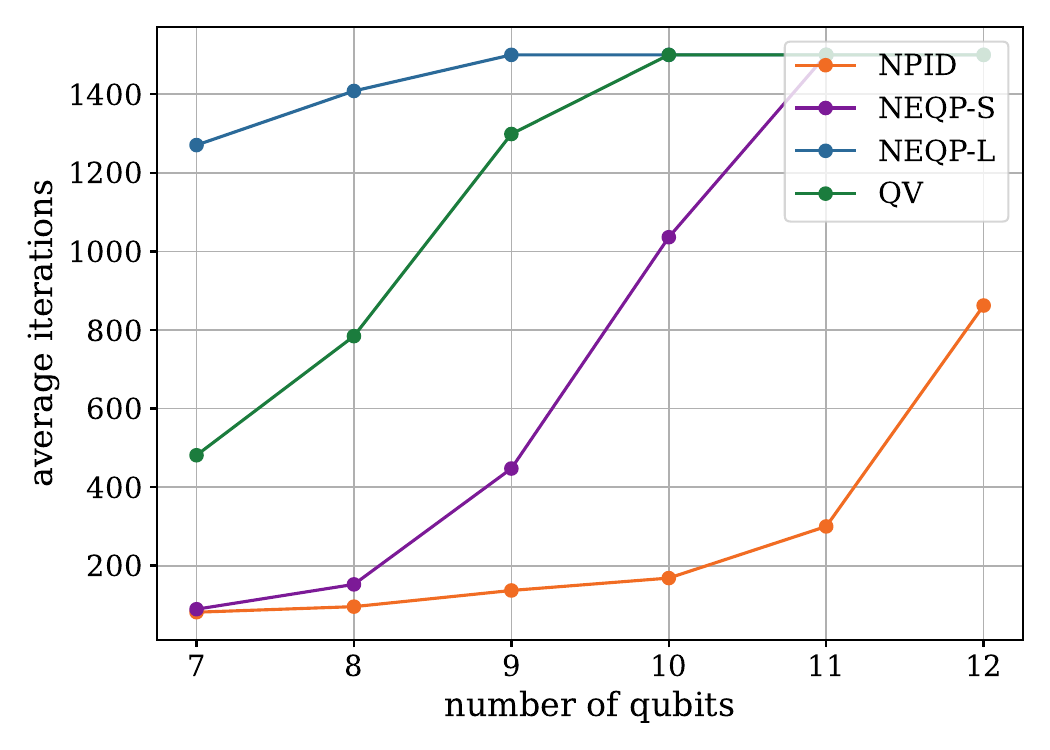}
    \caption{Number of iterations needed for the target cost function for different numbers of qubits.}
    \label{fig: avgsteps}
    \vspace{-15pt}
\end{figure}

Figure \ref{fig: avgsteps} illustrates the average number of iterations required by each model across different qubit counts. With the maximum iteration limit set to 1500, both QV and NEQP-L reached this upper bound once the number of qubits exceeded nine, while NEQP-S performed slightly better but still grew rapidly. In contrast, the NPID model showed superior performance: although its iteration count increased slightly with the number of qubits, its overall convergence speed remained significantly higher than that of NEQP-S, NEQP-L, and QV. Table \ref{tab: model_qubits} summarizes the average iteration count and convergence efficiency for each configuration. The average convergence efficiency $\hat{E_c}$ is computed as $avg(\sum\frac{max(Iteration)}{conv(Iteration)})$, where $max(Iteration)$ denotes the maximum number of iterations and $conv(Iteration)$ denotes the iterations required for convergence. As observed, NPID achieves an average convergence efficiency of 11.908, while the corresonding values for NEQP-S, NEQP-L, and QV are 5.544, 1.249, and 1.837, respectively. This demonstrates that NPID converges approximately 2–9 times more efficiently than the other models, achieving faster convergence and effectively mitigating the barren plateau.
\begin{table*}[htbp]
\centering
\caption{Average Number of iterations required to reach the target cost function value for different numbers of qubits.}
\begin{tabular}{cccccccc}
\hline
\textbf{Model} & \textbf{7 Qubits} & \textbf{8 Qubits} & \textbf{9 Qubits} & \textbf{10 Qubits} & \textbf{11 Qubits} &\textbf{12 Qubits} & \textbf{$\hat{E_c}$} \\ 
\hline
NPID & 82 & 96 & 138 & 177 & 328 & 881 & 11.908\\ 
NEQP-S & 90 & 153 & 448 & 1036 & 1500 & 1500 & 5.544 \\
NEQP-L  & 1270 & 1408 & 1500 & 1500 & 1500 & 1500 & 1.249 \\ 
QV  & 481 & 785 & 1299 & 1500 & 1500 & 1500 & 1.837 \\ 
\hline
\end{tabular}
\label{tab: model_qubits}
\vspace{-15pt}
\end{table*}

\begin{figure}
    \centering
    \includegraphics[width=0.8\linewidth]{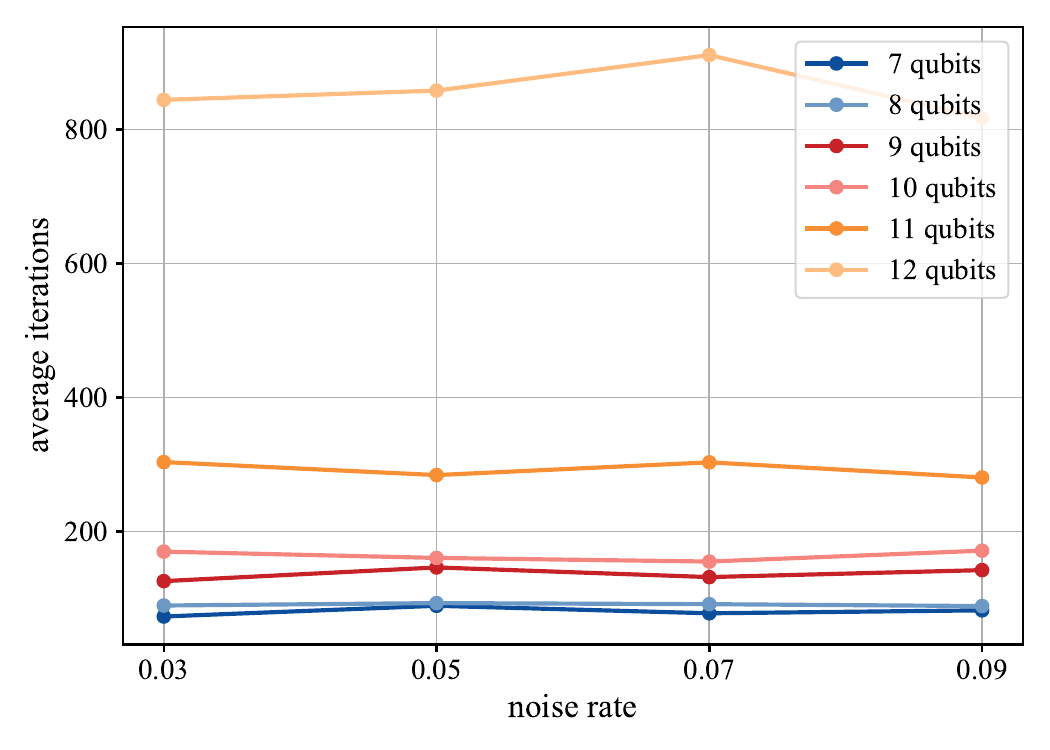}
    \caption{Average number of iterations required for NPID under different noise rates.}
    \label{fig: noiseiter}
    \vspace{-10pt}
\end{figure}

Table \ref{tab: noiseiter} compares the number of iterations required by the NPID model under different noise rates. Parameter noise was introduced at rates of 0.03, 0.05, 0.07, and 0.09, and the corresponding iteration counts for convergence were recorded for the same qubit configurations, as shown in Figure \ref{fig: noiseiter}. The results show that, for a fixed number of qubits, the iteration counts remain nearly consistent, fluctuating by only about 2–7\% (4.45\% on average), calculated as the ratio of standard deviation to mean. This indicates that, within a physically realistic range of parameter noise, variations in noise have little impact on the upper bound of the model’s convergence complexity. These findings further demonstrate the strong noise robustness of the proposed NPID model, suggesting its ability to effectively mitigate the barren plateau even in quantum circuits affected by parameter perturbations.

\begin{table*}[htbp]
\centering
\caption{Average Iterations comparison of NPID3 under varying qubit numbers and Noise rates.}
\begin{tabular}{ccccccc}
\hline
\textbf{Noise Rate} & \textbf{7 Qubits} & \textbf{8 Qubits} & \textbf{9 Qubits} & \textbf{10 Qubits} & \textbf{11 Qubits} &\textbf{12 Qubits} \\ 
\hline
0.03 & 74 & 90 & 127 & 171 & 304 & 845 \\ 
0.05  & 90 & 94 & 147 & 161 & 285 & 858 \\ 
0.07  & 79 & 92 & 133 & 156 & 304 & 911 \\ 
0.09 & 83 & 89 &  143 & 172 & 281 &  817 \\
\hline
\textbf{Fluctuation Rate} & 7.18\% & 2.10\% & 5.76\% & 4.08\% & 3.61\% &  3.98\%\\
\hline
\end{tabular}
\label{tab: noiseiter}
\vspace{-10pt}
\end{table*}

\vspace{-10pt}

\section{Summary}
\label{sec5}
\vspace{-10pt}

In this study, we propose an algorithm that incorporates a classical PID controller to update parameters in variational quantum circuits. To the best of our knowledge, this is the first work to introduce the concept of controllers from traditional control theory into variational quantum algorithms. We first outline the basic structure and workflow of a variational quantum algorithm, define the barren plateau phenomenon, and review several existing mitigation strategies, along with a brief construction of the PID controller. Next, we describe our proposed approach, which demonstrates universality by employing randomly generated qubit input states and randomly constructed quantum circuits. We then illustrate how the NPID model updates the circuit parameters during training. Simulation results show that the proposed NPID achieves faster and more stable convergence compared to NEQP and QV, without exhibiting significant fluctuations. Moreover, it maintains consistent performance across various noise levels, demonstrating strong robustness and scalability. This study bridges classical control theory and quantum optimization, offering a new perspective for enhancing the trainability of quantum neural networks in the NISQ era. Future work may explore extending this approach to other classical control strategies for mitigating the barren plateau in quantum machine learning with different types of noise.

    \vspace{-15pt}

\section*{Acknowledgments}
\vspace{-10pt}
This work was supported by a Graduate Research Assistantship at the AARC Lab, Department of Electrical and Computer Engineering, University of Alabama in Huntsville. The authors would like to thank Md Saiful Islam and Sophia Vanderwaal for some clarifications on issues related to control systems and physics, respectively.

\textbf{Code and Artifacts Availability:} All the artifacts and data used to produce this research will be made available at {\small\texttt{\kern-0.05em\url{https://github.com/AARC-lab/L4DC26_VQA}}} upon acceptance of the manuscript.

\bibliography{bib}
\appendix
\renewcommand{\theequation}{A.\arabic{equation}}
\setcounter{equation}{0}

\section{Proof of linearity of quantum circuit}\label{secA1}
We begin by briefly reviewing a fundamental concept from quantum mechanics. In quantum theory, any observable or transformation physical quantity $O_p$ can be represented as a linear operator acting on a Hilbert space. For any two quantum states $\ket{\psi_a}$ and $\ket{\psi_b}$, and for any complex scalars $a$ and $b$, we have
\begin{equation}
    O_p(a\cdot \ket\psi_a + b\cdot \ket{\psi_b}) = a\cdot O_p\ket{\psi_a} + b \cdot O_p\ket{\psi_b}
\end{equation}

For a quantum circuit represented by the unitary operator $U$, the entire transformation can be expressed as
\begin{equation}
    \ket{\psi_{out}} = U\ket{\psi_{in}}
\end{equation}
Therefore, for any two quantum states $\ket{\psi_m}$ and $\ket{\psi_n}$, and for any complex scalars $m$ and $n$, we have
\begin{equation}
    U(m\cdot \ket\psi_m + n\cdot \ket{\psi_n}) = m\cdot U\ket{\psi_m} + n \cdot U\ket{\psi_n}
\end{equation}
Although the unitary operator $U(\theta)$ is generally nonlinear in its parameters $\theta$, under a small-perturbation approximation $\Delta \theta \rightarrow 0$, it can be locally linearized as
\begin{equation}
    \begin{aligned}
        U(\theta) &= e^{-i\theta H} \\
        U(\theta + \Delta \theta)  &= e^{-i(\theta+\Delta\theta)H} \\
        &=e^{-i\theta H}\cdot e^{-i\Delta\theta H} \\
        &\approx e^{-i\theta H} \cdot(I-i\Delta\theta H)\\
        & = U(\theta)-i\Delta\theta H U(\theta)
    \end{aligned}
\end{equation}
where $H$ denotes the corresponding Hermitian matrix.
This local linearity justifies treating the parameterized quantum circuit as a linear system for the purpose of applying PID control theory.

\end{document}